\documentclass[fleqn,usenatbib]{mnras}

\usepackage{newtxtext,newtxmath}
\usepackage[T1]{fontenc}

\DeclareRobustCommand{\VAN}[3]{#2}
\let\VANthebibliography\thebibliography
\def\thebibliography{\DeclareRobustCommand{\VAN}[3]{##3}\VANthebibliography}

\usepackage{graphicx}	% Including figure files
\usepackage{amsmath}	% Advanced maths commands
\usepackage{amssymb}	% Extra maths symbols

\title[SB2s in LAMOST-MRS]{Detection of 12426 SB2 candidates in the LAMOST-MRS, using a binary spectral model.}

\author[M. Kovalev et al.]{
Mikhail Kovalev,$^{1,2,3}$\thanks{E-mail: mikhail.kovalev@ynao.ac.cn}
Zenghua Zhou,$^{1,2}$
Xuefei Chen,$^{1,2,4}$
Zhanwen Han$^{1,2,4}$
\\
$^{1}$Yunnan Observatories, China Academy of Sciences, Kunming 650216, China\\
$^{2}$Key Laboratory for the Structure and Evolution of Celestial Objects, Chinese Academy of Sciences, Kunming 650011, China\\
$^{3}$Sternberg Astronomical Institute, Leninskie Gory, Moscow 119992, Russia\\
$^{4}$Center for Astronomical Mega-Science, Chinese Academy of Sciences, 20A Datun Road, Chaoyang District, Beijing 100012, China\\
}

\date{Accepted 2023-10-16. Received 2023-10-15; in original form 2023-09-11}

\pubyear{2023}

\def\kms{\,{\rm km}\,{\rm s}^{-1}}

\def\feh{\hbox{[Fe/H]}}

\newcommand{\teff}{{T_{\rm eff}}}
\newcommand{\rv}{{\rm RV}}
\def\Vmic{V_{\rm mic}}

\def\vsini{V \sin{i}}%\def\vsini{V{\rm \sin }i}
\def\logg{\log{\rm (g)}}
\def\snr{\hbox{S/N}}
\def\drv{\hbox{|$\Delta$ RV|}}
\def\imp{f_{\rm imp}}
\newcommand{\ha}{\hbox{H$\alpha$}}

\begin{document}
\label{firstpage}
\pagerange{\pageref{firstpage}--\pageref{lastpage}}
\maketitle

\begin{abstract}%It can reliably detect SB2 candidates for double-lined binaries with $\vsini_1+\vsini_2<300\kms$ if the radial velocity separation is large enough. 
We use an updated method for the detection of double-lined spectroscopic binaries (SB2) using $\vsini$ values from spectral fits. The method is applied to all spectra from LAMOST MRS. Using this method, we detect 12426 SB2 candidates, where 4321 are already known and 8105 are new discoveries. We check their spectra manually to minimise possible false positives. We also detect several cases of contamination of the spectra by solar light. Additionally, for candidates with multiple observations we compute mass ratios with systemic velocities and determine Keplerian orbits. We present an updated catalogue of all SB2 candidates together with additional information for some of them in separate data tables. 
%No references should appear in the abstract.
\end{abstract}


%%%%%%%%%%%%%%%%%%%%%%%%%%%%%%%%%%%%%%%%%%%%%%%%%%

%%%%%%%%%%%%%%%%% BODY OF PAPER %%%%%%%%%%%%%%%%%%

\section{Introduction}

\cite{cat22} presented a new method for detection of the double-lined spectroscopic binaries (SB2s) in LAMOST (Large Sky Area Multi-Object fiber Spectroscopic Telescope, also known as Guoshoujing telescope) Medium Resolution Survey (MRS). It is based on a clear correlation between the radial velocity separation ($\drv=|\rv_1-\rv_2|$) estimated by the binary model and the value of the projected rotational velocity $\vsini_0$, derived by the single-star model for SB2 systems. It was proven that this method is able to identify SB2s even when the standard analysis of cross-correlation functions (CCF) fails. Two other studies, \cite{tyc,j0647}, presented detailed analyses of the two SB2 systems using multiple spectroscopic observations simultaneously and proved that it is reliable. A recent study, \cite{satcon} explored the problem of possible contamination of LAMOST-MRS spectra by the solar light reflected by the nearby objects and found out, that some SB2 candidates can be result of such contamination.  
\par
In this paper we continue the analysis of the LAMOST MRS using a binary spectral model. The methods from \cite{cat22} are updated and applied to all available spectra. Thus we significantly expand our catalogue, which now has six times more SB2 candidates, and also present more detailed results for many of them. %\cite{tyc,j0647} successfully analysed multiple spectroscopic observations simultaneously for two interesting binary systems and now we apply the same methods to a larger dataset.

\par
The paper is organised as follows: in Sections~\ref{sec:obs} and \ref{sec:methods}, we describe the observations and methods. Section~\ref{results} presents our results. In Section~\ref{discus} we discuss the results. In Section~\ref{concl} we summarise the paper and draw conclusions.

\section{Observations}
\label{sec:obs}
%\subsection{Spectra}
LAMOST is a 4-metre quasi-meridian reflective Schmidt telescope with 4000 fibers installed on its $5\degr$ FoV focal plane. These configurations allow it to observe spectra for at most 4000 celestial objects simultaneously (\cite{2012RAA....12.1197C, 2012RAA....12..723Z}).
 In comparison with \cite{cat22} which was restricted only to time-domain spectra, in this paper we downloaded all available spectra from \url{www.lamost.org}, including non-time-domain (NT) spectra.	We use the spectra taken at a resolving power of $R=\lambda/ \Delta \lambda \sim 7\,500$. Each spectrum is divided on two arms: blue from 4950\,\AA~to 5350\,\AA~and red from 6300\,\AA~to 6800\,\AA.~We convert the heliocentric wavelength scale in the observed spectra from vacuum to air using \texttt{PyAstronomy} \citep{pya}. Observations are carried out in MJD=58407.6-59736.8 days, spanning an interval of 1329 days.
 We selected only spectra stacked  within a single night\footnote{Each exposure contains a sequence of the several short 20 min individual exposures, which were stacked to increase $\snr$.} and apply a cut on the signal-to-noise ($\snr$). In total we have $1\,805\,736$ spectra from $840\,826$ targets, where the $\snr\geq20$ ${\rm pix}^{-1}$ in any of spectral arms. The majority of the spectra (one half) sample the $\snr$ in the range of 40-250 ${\rm pix}^{-1}$. The number of exposures varies from 1 to 40 per target, as noisy exposures were not selected for many targets.

\section{Methods} 
\label{sec:methods} 
\subsection{Spectroscopic analysis}%maybe move to the appendix? %in Appendix~\ref{sec:payne}. 
\label{sec:specfit}

We use the same spectroscopic models and method as \cite{tyc} to analyse individual LAMOST-MRS spectra, see brief description below.
The normalised binary model spectrum is generated as a sum of the two Doppler-shifted normalised single-star spectral models ${f}_{\lambda,i}$\footnote{they are designed as a good representation of the LAMOST-MRS spectra, see Appendix~\ref{sec:payne} for details}, scaled according to the difference in luminosity, which is a function of the $\teff$ and stellar size. We assume both components to be spherical.% and use the following equation:    

The binary model spectrum is later multiplied by the normalisation function, which is a linear combination of the first four Chebyshev polynomials \citep[similar to][]{Kovalev19}, defined separately for the blue and red arms of the spectrum. The resulting spectrum is compared with the observed one using \texttt{scipy.optimise.curve\_fit} function, which provides the optimal spectral parameters and radial velocities (RV) for each component plus the stellar radii ratio $k_R={R_1}/{R_2}$ and two sets of four coefficients of the Chebyshev polynomials. We keep the metallicity equal for both components. In total we have 18 free parameters for a binary fit. We estimate the goodness of the fit parameter by reduced $\chi^2$.

\par 
Additionally, every spectrum is analysed by a single star model, which is identical to a binary model when both components have all equal parameters, so we fit only for 13 free parameters. Using this single star solution\footnote{throughout paper it has subscript ``0" or ``single"} we compute the difference in reduced $\chi^2$ between two solutions and the improvement factor ($\imp$), computed using Equation~\ref{eqn:f_imp} similar to \cite{bardy2018}. This improvement factor estimates the absolute value difference between two fits and weights it by the difference between the two solutions.

\begin{align}
\label{eqn:f_imp}
f_{{\rm imp}}=\frac{\sum_{\lambda=\lambda_{min}}^{\lambda=\lambda_{max}}\left[ \left(\left|{f}_{\lambda,{\rm single}}-{f}_{\lambda}\right|-\left|{f}_{\lambda,{\rm binary}}-{f}_{\lambda}\right|\right)/{\sigma}_{\lambda}\right] }{\sum_{\lambda=\lambda_{min}}^{\lambda=\lambda_{max}}\left[ \left|{f}_{\lambda,{\rm single}}-{f}_{\lambda,{\rm binary}}\right|/{\sigma}_{\lambda}\right] },
\end{align}
where ${f}_{\lambda}$ and ${\sigma}_{\lambda}$ are the observed flux and corresponding uncertainty, ${f}_{\lambda,{\rm single}}$ and ${f}_{\lambda,{\rm binary}}$ are the best-fit single-star and binary model spectra, and the sum is over all wavelength pixels.% This improvement factor allowed us to select exposure where binary model has maximal performance

%\par
% The current paper focuses on SB2 identification of a single epoch spectrum, and we will present the results of the simultaneous analysis of the multiple spectra in our future paper (Kovalev et al. in prep).  

\subsection{Selection of SB2 candidates}
\label{sec:selection}
Here we use the same logic as \cite{cat22} for selection: we separate single stars from binary solutions. If we try to fit the double-lined spectrum with $\drv>0$ using a single star model there are three possible outcomes:
\begin{enumerate}
    \item none of the spectral components are well constrained, but $\vsini_0$ proportional to $\drv$ is large enough that the broadened model covers both of them;
    \label{case1}
    \item only one spectral component (usually the primary) is well fitted, while the other is completely ignored by the single-star spectral model;
    \label{case3}
    \item the spectrum is poorly fitted because the synthetic model is unable to reproduce a real spectrum due to missing physics (usually emission lines, molecular bands etc.) or data processing artifacts.%, see Appendix~\ref{sec:bads}.
    \label{case2}
\end{enumerate}
%(see Appendix~\ref{sec:payne} for details)
For all these cases binary model fits the spectrum much better and $\imp$ is large. However case~\ref{case1} is most useful to select SB2s. Even small $\drv$, which is insufficient to split spectral lines due to finite spectrograph's resolution,  will cause a change to the spectral lines: they will become broader and the fitted value of $\vsini_0$ will increase in comparison with the value from the spectrum of that binary taken when $\drv=0$. Case~\ref{case3} can be excluded by requiring $\rv_{1}$ to differ from $\rv_0$, as for such a case value $\vsini_0$ is useless for SB2 identification. However, it can be used if the binary model had a good fit, so we put some thresholds on $\imp$ and the ratio of $\chi_{\rm single}/\chi_{\rm binary}$. Multiple epochs can also allow us to find an SB2 candidate by $\vsini_0$ variation, which it is proportional to $\drv$, although such a variation can be caused by a real change of $\vsini$. %if double-lined structure unresolved in the spectrum.    

\par
Now we need to separate out the single stars. When a single star spectrum with $\vsini_0$ is fitted by a binary model there are three possible outcomes:
\begin{enumerate}
    \item the spectrum is well fitted by a twin binary model, consisting of two components similar to the real star with small $\drv\sim0$,  therefore $\vsini_{1,2}\sim \vsini_0$ and $\vsini_1+\vsini_2\sim 2\vsini_0$;
    \label{case01}
    \item the spectrum is well fitted by the primary component, which is almost identical to the real star ($\vsini_1\sim\vsini_0$), and a small or negligible contribution from the secondary component, which can have any value $\vsini_2$, although usually it is quite large, so the secondary spectrum looks completely ``flat"; 
    \label{case02}
    \item the spectrum is poorly fitted as the synthetic model is unable to reproduce the real spectrum. In this case, the binary model is desperately trying to compensate for the missing spectral information by combining two spectral components, usually with significant improvement relative to best single star model.%  see Appendix~\ref{sec:bads}.
    \label{case03}
\end{enumerate}% However if we fit SB2 spectrum with binary model even small radial velocity separation will cause change spectral lines: they will become broader and fitted value of  will increase. It is not surprising as rotational broadening is also caused by the Doppler shift of the light wavelength coming from the stellar surface.
For cases~\ref{case01} and \ref{case02}, $\imp$ is usually small, but for case~\ref{case03} it can be quite large.
Thus we can select SB2 candidates with $\drv>0$ by selecting the spectra with $\vsini_1+\vsini_2 +\vsini_{\rm min}< \vsini_0$ and $\rv_1\neq\rv_0$, while ``bad fits" from case~\ref{case03} can be excluded using a cut on $\imp$. $\vsini_{\rm min}$ takes into account the possible uncertainties in the $\vsini$ measurements. However we should note that these criteria will not select spectra of fast rotators, as in this case $\vsini_1+\vsini_2>\vsini_0$ for small $\drv>0$. They can be selected only for a significant $\drv$. This is our standard selection. %One can find various fit examples in Appendix~\ref{fit_examples}.%If multiple epochs are available we can find SB2 candidate by $\vsini_0$ variation (Kovalev et al. in prep.). 
More SB2 candidates from case \ref{case3} with the single-star model fitting only a primary are selected using $\rv_1\sim\rv_0$ and cuts on $\imp$ and $\chi_{\rm single}/\chi_{\rm binary}$, which can be found empirically. We call such selection ``selected by primary". An interesting system J065032.45+231030.2 mentioned, but not selected in \cite{cat22}, is selected by these criteria. 
%\par

\subsection{Updates relative to previous version}
In this section we summarize all updates applied to the methods in this paper, in comparison with \cite{cat22}:
\begin{enumerate}
    \item we switch from neural network based spectral model of the single star to simple linear interpolation, see Appendix~\ref{sec:payne},
    \item  for the binary spectral model we directly use the ratio of stellar radii $k_R$ as a fitting parameter, instead of the mass ratio which was later combined with the difference of the surface gravities $\logg$ to get the ratio of stellar radii.
    \item we introduce a new ``selection by primary" for SB2 candidates, where the single-star spectral model ignores  secondary component.
    \item we manually checked all selected SB2 candidates to minimise number of false-positives detections. 
\end{enumerate}

\section{Results}
\label{results}

\begin{figure}%[!htb]
	\includegraphics[width=\columnwidth]{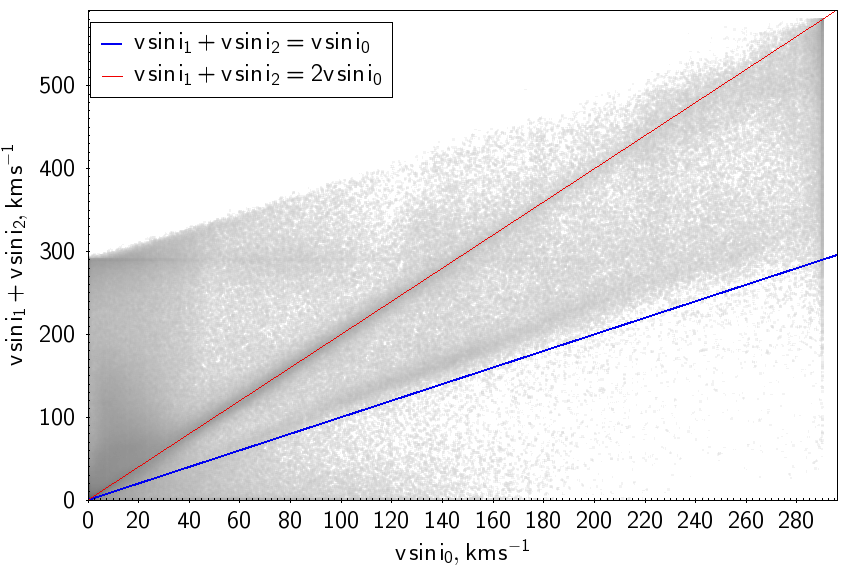}
	\includegraphics[width=\columnwidth]{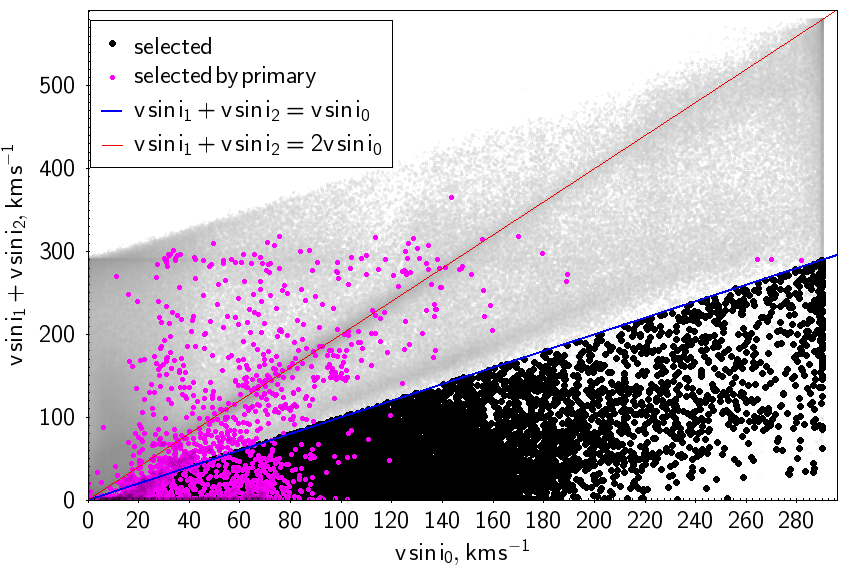}
    \caption{Values of $\vsini$ for the observed dataset (top panel) and selected spectra (bottom panel). Solid lines show the functions $\vsini_1+\vsini_2=\vsini_0$ (blue) and $\vsini_1+\vsini_2= 2\vsini_0$ (red). Selected stars confirmed via visual inspection are shown with black circles (standard selection) and pink circles (selected by primary). }
    \label{fig:select}
\end{figure}

\subsection{Quality cuts}
We carefully check the quality of the spectral fits through visual inspection of the plots. Our LAMOST-MRS dataset contains spectra from various targets and some of them can be poorly fitted by our spectral model (red super giants, very hot stars etc.). Therefore we introduce several quality cuts on the fitted parameters, see Table~\ref{tab:cuts}. These cuts keep $1\,497\,275$ spectra (83 per cent) from the original data set. Additionally we find $1367$ spectra, where the wavelength scale is clearly shifted between the blue and left arms, which is six times more than in \cite{cat22}.%see Appendix~\ref{sec:bads}.    

\begin{table}%\small
    \centering
    \caption{Quality cuts and selection criteria.}
    \begin{tabular}{l}
\hline
\hline
bad fits - any of the following cuts (293203 spectra):\\
\hline
$\chi^2_{\rm single}$  >12000\\
$\drv$  >1320 $\kms$\\
|$\rv_0$|  >400 $\kms$\\
|$\log{k_R}$|  >0.99\\
$k_{5000\AA}/(1+k_{5000\AA})$  >0.96\\
$\feh_0$<-0.77 and $\teff_0$>8600 K\\
$\feh_1$<-0.77 and $\teff_1$>8600 K\\
|$\feh_0$|>0.49 and $\vsini_0$>289 $\kms$\\
|$\logg_0-3$|>1.89 and $\vsini_0$>289 $\kms$\\
|$\logg_0-3$|>1.89 and ${\teff}_0$>8690 K\\
|$\logg_1-3$|>1.89 and ${\teff}_1$>8690 K\\
Bad $\lambda$ calibration in 1367 spectra\\
\hline
standard SB2 selection (21962 spectra):\\% - all criteria should match\\
not in bad fits\\
$\imp$  >0.10\\
|$\rv_1-\rv_0$|  >10 $\kms$\\
$\vsini_1+\vsini_2$ +0.5< $\vsini_0$\\
\hline
SB2 selection by primary (2613 spectra):\\% - all criteria should match\\
not in bad fits\\
$\imp$  >0.20\\
|$\rv_1-\rv_0$|  <=10 $\kms$\\
$\chi_{\rm single}/\chi_{\rm binary}>1.5$\\
\hline

    \end{tabular}
    \label{tab:cuts}
\end{table}

\subsection{Selected SB2 candidates}%add Table available in electronic version selection of SB2 candidates using $\vsini$ on the bottom
%Solid lines are showing functions $\vsini_1+\vsini_2=\vsini_0$ (blue) and $\vsini_1+\vsini_2= 2\vsini_0$ (red). 
 We show the plot with $\vsini$ values, like in \cite{cat22}, in the top panel of Fig.~\ref{fig:select}. It is clearly seen that many datapoints follow the red line and a slightly broad region above the blue line, which are the places where single stars should be. We have overdensities at $\vsini_0=290\,\kms$ and $\vsini_1+\vsini_2=290\,\kms$, where some value of $\vsini$ is maximal. The first one is due to a restriction on extrapolation, while the second is formed by the stars where one of the components have maximal value of $\vsini$ and usually shows no spectral lines. In comparison with \citet{cat22} there is no gap at $\vsini_0\sim90\,\kms$. The gap was caused by a systematic problem in the fitting mechanism, which is gone as we replaced the neural network based spectrum generator by the simple linear interpolation. On the bottom panel we show selected SB2s candidates, confirmed by the manual inspection of the plots. We keep 19503 spectra in standard selection and 1839 in selected by primary subsets. In total we select 21342 spectra of 12426 stars.

\subsection{Mass ratio measurements using the Wilson method.}
\label{sec:multi}

For SB2 candidates with several ($>3$) spectra and good $\rv_{1,2}$ measurements we can get dynamical mass ratios using the Wilson plot \citep{wilson,kounkel}.
If two components in our binary system are gravitationally bound, their radial velocities should agree with the following equation \citep{wilson}:  

\begin{align}
\label{eqn:asgn}
    {\rm RV_{1}}=\gamma_{\rm dyn} (1+Q_{\rm dyn}) - Q_{\rm dyn} {\rm RV_{2}},
\end{align}
where $Q_{\rm dyn}={M_1}/{M_{2}}$ - mass ratio of binary components and $\gamma_{\rm dyn}$ - systemic velocity. Using this equation we can directly measure the systemic velocity and mass ratio by fitting the line, using orthogonal distance regression (ODR) method \citep{odr}\footnote{ \url{https://docs.scipy.org/doc/scipy/reference/odr.html}} which can handle uncertainty in both variables.
\par
\begin{figure}%[!htb]
	\includegraphics[width=\columnwidth]{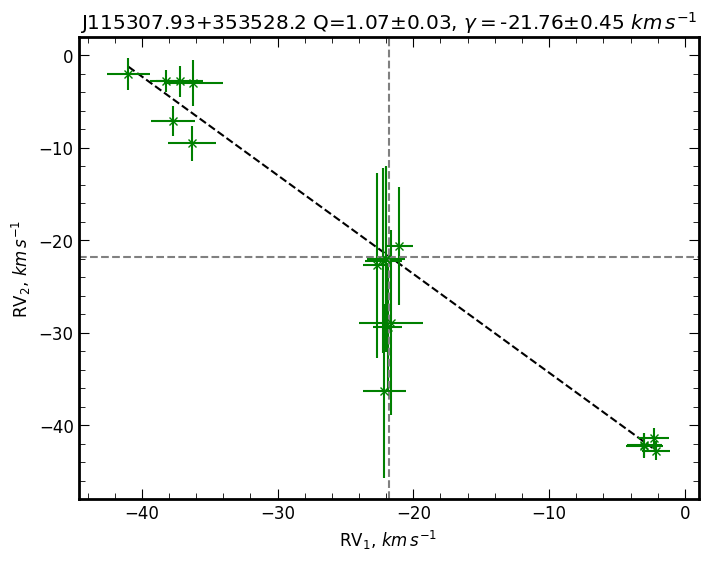}
    \caption{Wilson plot for J115307.93+353528.2 SB2 candidate. The mass ratio and systemic velocity from the ODR fit are shown in the title. The $\gamma$ is also shown by dashed vertical and horizontal lines.}
    \label{fig:wilson}
\end{figure}

In Fig.~\ref{fig:wilson} we show example of such fitting in the Wilson plot. Note that many datapoints with large errors are concentrated close to $\gamma$. Fortunately they did not affect final result as we have enough measurements near the extreme $\rv$ positions.   

We present the results for dynamical mass ratios and $\gamma$ in Table~\ref{tab:wilson}. %The results for two stars analysed in \cite{tyc,j0647} agree very well with previous estimates: $Q_{\rm dyn}=$ $Q_{\rm lit}=$ and $Q_{\rm dyn}=$ $Q_{\rm lit}=$. 
Also we computed weighted the Pearson correlation coefficient for all datapoints to select good solutions that follow the straight line.

\begin{align}
    R_{w}=\frac{cov(x,y)}{\sqrt{cov(x,x) cov(y,y)}}\\
    cov(x,y)=\sum \frac{w(x-\langle x\rangle)(y-\langle y\rangle) }{\sum w}\\
    \langle x\rangle=\frac{\sum w x}{\sum w}, w=\sigma^{-1}
\end{align}

Some values of mass ratios are very small and even negative (which have no physical meaning). \cite{kounkel} also found such values using similar method. We suspect that such values are from spectroscopic triples or from chance alignment of spectroscopic binary with another star. We study three these objects in separate paper \citep{threeSB2s}.%need to put citation on arxive paper later

\subsection{Orbital solutions}
\begin{figure}%[!htb]
	\includegraphics[width=\columnwidth]{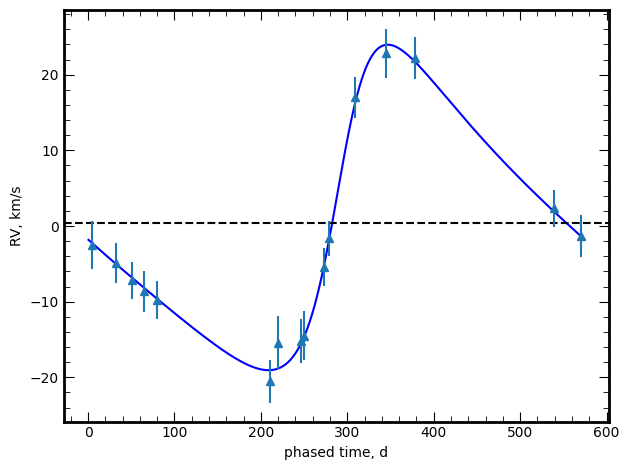}
    \caption{Example of the Keplerian orbit for primary component of J065032.45+231030.2 derived using {\sc GLS}.}
    \label{fig:orbfit}
\end{figure}

For stars with $N\geq 6$, good $\rv_{1}$ measurements, and $R_w<-0.95$, we use a Generalised Lomb-Scargle periodogram ({\sc GLS}) by \cite{gls} to get orbits using circular and Keplerian solutions. We checked periods in the range $P=1,1000$ d on the regular grid of $e=0,0.8$ and $\omega=0,359^\circ$ with steps 0.01 and $1^\circ$ respectively. All parameters have errors provided by GLS, except for Keplerian orbits for $\rv$ time series with six measurements, which have infinite errors. For several solutions with $P=1000$ d we refined the solution using periods in the range $P=100,10000$ d. The example of the Keplerian orbit for J065032.45+231030.2 is shown on Figure~\ref{fig:orbfit}.

\begin{figure}%[!htb]
	\includegraphics[width=\columnwidth]{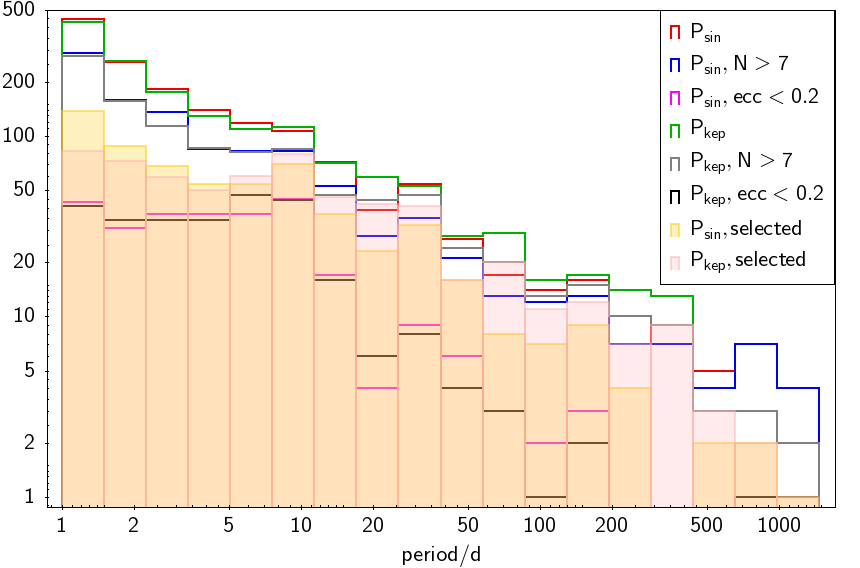}
    \caption{Histogram for periods derived using {\sc GLS}.}
    \label{fig:histp}
\end{figure}
We show histograms for the periods computed with circular and Keplerian orbits in Figure~\ref{fig:histp}.
A significant fraction of the stars with $P<2$ d have large eccentricities ($e>0.2$) which is suspicious, as close binaries usually have circular orbits \citep{kounkel}. This is likely due to the use of stacked spectra for $\rv$ determination with small temporal resolution. Thus we decided to check and discard many such suspicious solutions. 
After visual inspection of the results we selected 616 reliable solutions both circular and Keplerian, listed in Table~\ref{tab:orbits}. It is interesting that for J064726.39+223431.7 even using stacked spectra we were able to get a period $P=1.217842\pm0.000012$ d which is very close to results from analysis of individual short exposures  $P=1.217770 \pm 0.000003$ d \citep{j0647}. We think this is due to large semiamplitude ($K\sim 122~\kms$) of the orbit in this system. %For J091914.65+423348.0 $P_{\rm sin}=7.0573\pm0.0006$  $P=7.0565 \pm 0.0004$ d \citep{tyc}.

\section{Discussion}
\label{discus}

\subsection{Comparison with other SB2 catalogues}%We check these matches and confirm that they are SB2 candidates, see Fig.\ref{fig:elbadry}.
We make cross matches with several available SB2 catalogues, and find more matches than \cite{cat22}, which is not surprising as we analysed significantly more targets. We find 863 matches with SIMBAD database \citep[][]{simbad}, labelled as SB*, seven matches with SB9 \citep[][]{sb9},  141 matches with \cite{traven20}, and 367 matches with \cite{kounkel} catalogues. Cross matching with \cite{bardy2018} we find 24 stars listed in their SB2 table. Surprisingly, we find four matches (three new in comparison with \cite{cat22}) with their single star table and one match with the SB1 table.  We find no matches with either the SB3 table or the SB2 table with unseen third component. Among LAMOST-MRS based papers we find 1276 SB2 matches and 50 SB3 matches with \citet{li2021}, 292 matches with \citet{songK2}, 1144 matches with \citet{zhangbo22}\footnote{among their final sample of 2198 SB2 candidates} and 436 matches with \cite{Zheng_2023}. We find 594 matches with recent Gaia DR3 non-single stars orbits catalogue\citep[][]{gaia_dr3multiple}, although only 236 and 231 are labelled as SB2 and SB1 respectively. 
By comparing with catalogue in \cite{cat22} we confirm 2073 out of 2460 SB2 candidates, with rest unable to pass visual inspection of the plots.
\par
In total our catalogue includes 12426 SB2 candidates, where 4321 are known and 8105 are new. Some of them are definitely spectroscopic triples or even quadruples, however the current selection method cannot distinguish them from SB2s. We note that some SB2 candidates can be chance alignments due to relatively large fiber diameter of the spectrograph (3\arcsec) or a result of contamination of the spectra by solar light, reflected by close objects \citep{satcon}, see Section~\ref{cont}. %Our sample is limited to time-domain spectra with $\snr>25$, therefore the number of matches with other LAMOST-MRS based studies is not very high. We plan to extend it to non time-domain spectra in our future paper.

\par
\begin{figure}%[!htb]
	\includegraphics[width=\columnwidth]{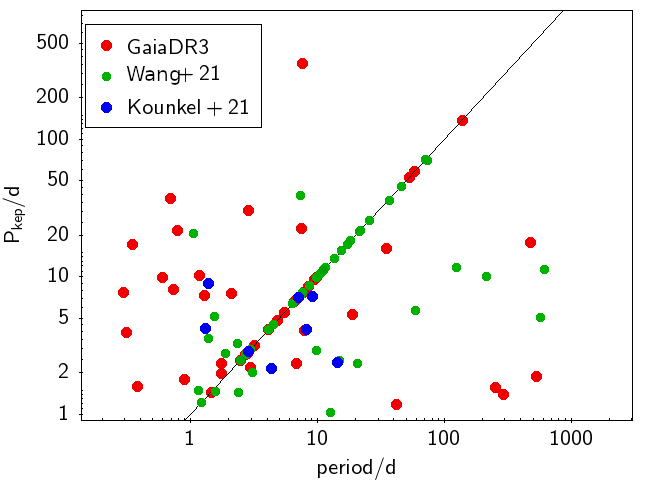}
    \caption{Comparison of the Keplerian periods derived using {\sc GLS} with literature values.}
    \label{fig:percom}
\end{figure}

We compare our {\sc GLS} periods with available periods from the literature in Figure~\ref{fig:percom}. We found 44, 48 and 8 period values in \cite{gaia_dr3multiple},\cite{songK2} and \cite{kounkel}. Many values agree, but a significant fraction are very different, so we recommend to use our periods with care. Currently we use only the primary $\rv$ to determine the orbit, but we plan to use data for both components \citep[similarly to ][]{kounkel} in our future study.

\subsection{Contaminated spectra}
\label{cont}%2019-02-17UTC12:51:00
In \cite{cat22} the spectrum of J055923.95+303104.2 taken at MJD=58531.535 d was indicated as a possible SB2 candidate, where a single star model fits only narrow lines of the secondary component. This narrow-lined component was absent in all other spectral observations, thus we proposed that this spectrum was taken during the partial eclipse. However there is another simple explanation: this is scattered light from the full Moon located at $\sim29\degr$ from the field center, which can be seen as narrow-lined component with $\rv_2\sim-\rv_{\rm heliocentric}$.
\par
\begin{figure*}%[!htb]
	\includegraphics[width=0.83\textwidth]{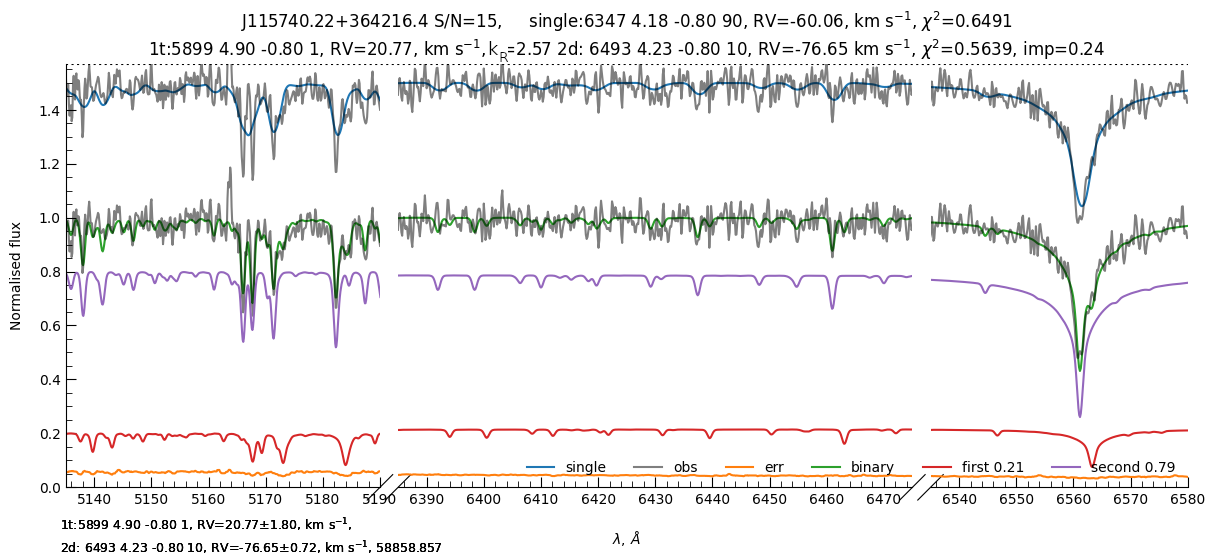}
	\includegraphics[width=0.83\textwidth]{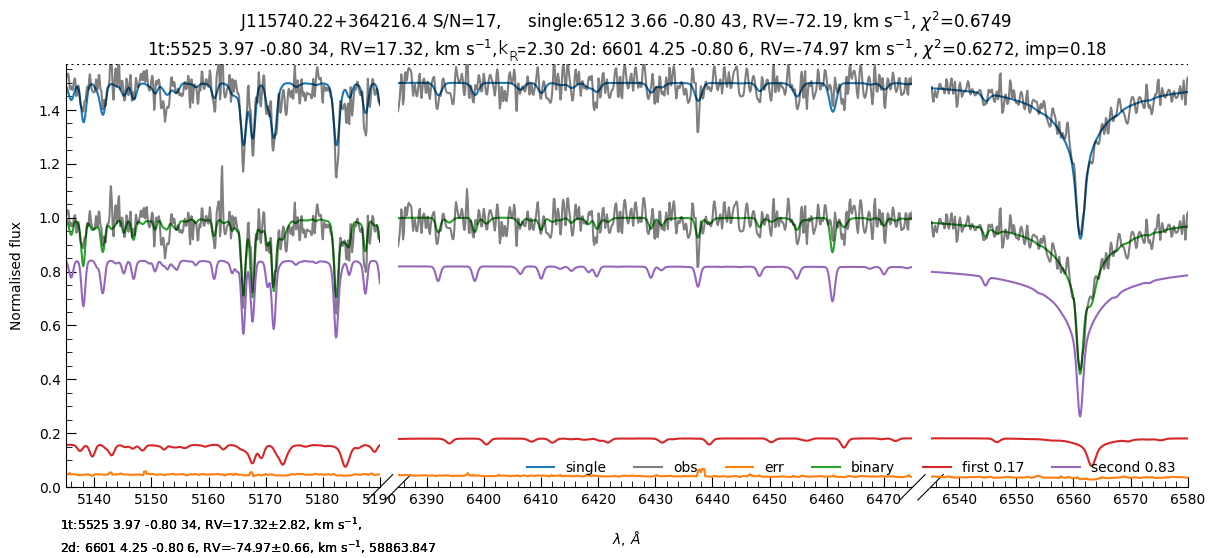}
    \includegraphics[width=0.83\textwidth]{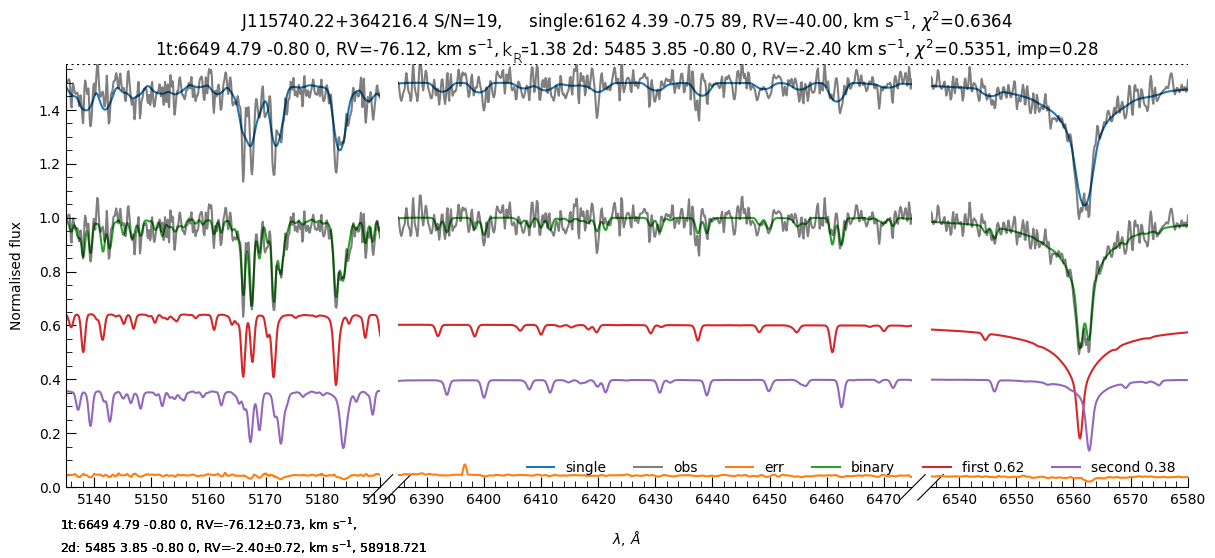}
    \caption{Examples of the spectra with possible contamination by the scattered Moon's light. The fitting by the binary spectral model and single-star model are shown as green and blue (with offset 0.5) lines respectively. The observed spectrum and its error are shown as a gray and orange lines respectively. The primary (magenta line) and secondary (red line) components are labelled as "second" and "first" with contribution to total light at $\lambda=5000$~\AA.~The spectral parameters ($\teff,~\logg,~\feh,~\vsini$) from the single star model fit and the binary model fit are shown in the titles.}
    \label{fig:moon}
\end{figure*}

We found several targets with similar contamination. See Figure~\ref{fig:moon} for J115740.22+364216.4. Three spectra taken at epochs MJD=$58858.857,\,58863.847,\,58918.721$ d have Moon's phases $14.78,\,18.78,\,14.92$ d (out of 29.53 d) respectively. The $\rv_{\rm heliocentric}=-20.37,\,-18.96,\,2.98~\kms$ from the FITS file headers, taken with opposite sign, agree very well with $\rv_2=20.77\pm1.80,\,17.32\pm2.82,\,-2.40\pm0.72~\kms$ derived by our binary model.   

\par
\begin{figure*}%[!htb]
	\includegraphics[width=0.83\textwidth]{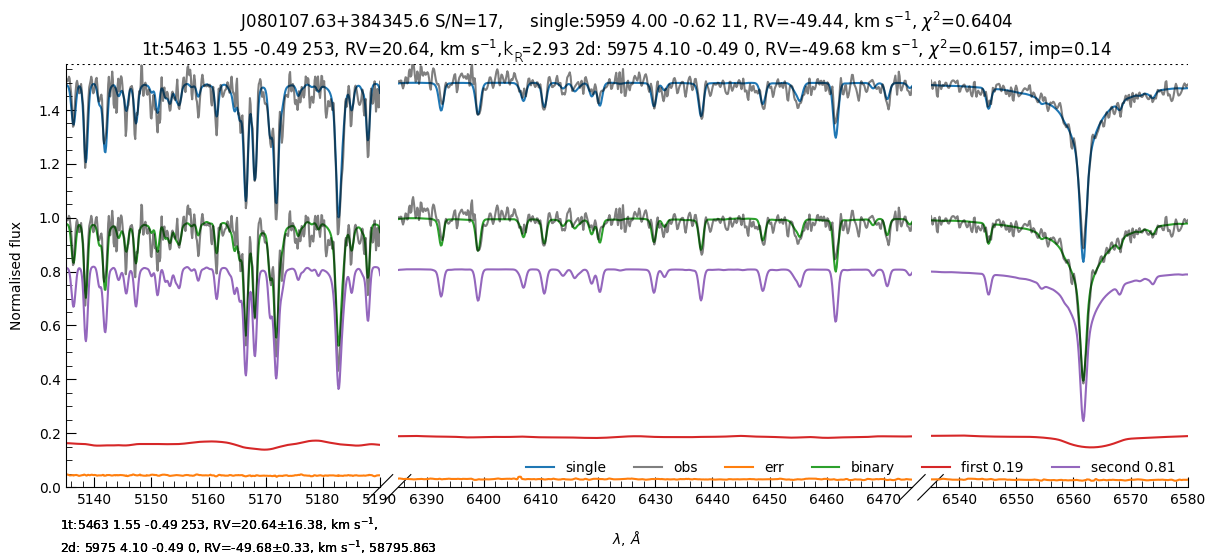}
	\includegraphics[width=0.83\textwidth]{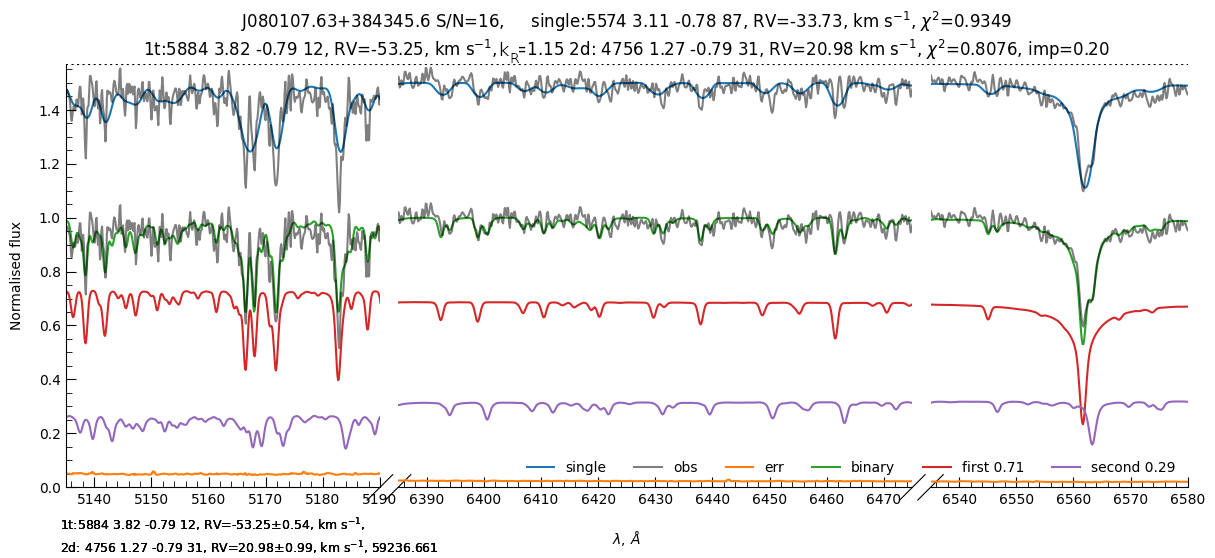}
    \caption{Same as Fig.\protect\ref{fig:moon} but for J080107.63+384345.6: the top panel shows the spectrum is well fit by the single-star model, and the spectrum in the bottom panel indicates an some unknown contamination. }% Observed spectrum and its error are shown as a gray and orange lines respectively. Primary (magenta line) and secondary (red line) components are labeled as "second" and "first" with contribution to total light at $\lambda=5000$~\AA.~Spectral parameters ($\teff,~\logg,~\feh,~\vsini$) from single star model fit and binary model fit are shown in the titles.}
    \label{fig:puzzle}
\end{figure*}

Another interesting target is J080107.63+384345.6 with $G=14.691$ mag. Only one ( observed at MJD=59236.661 d) of it's spectra  have been selected as SB2 candidate with significant $\Delta\rv\sim73~\kms$ between components, while all other seven are well fitted by single-star model with $\rv\sim-50~\kms$, see Figure~\ref{fig:puzzle}. The ODR fit provides very small $Q_{\rm dyn}~\sim0.05$ which is very suspicious, so we suspected contamination by reflected solar light from artificial satellites. However it is unlikely as contaminant's $\rv=20.98\pm0.99~\kms$ is not consistent with $\rv_{\rm heliocentric}=3.96~\kms$. The Moon's phase is also not full 9.62 d. We also checked for other solar-system bodies with Minor Planet Center checker\footnote{\url{https://minorplanetcenter.net/cgi-bin/checkmp.cgi}}, but found no objects brighter than $V=21$ mag close to our target at the moment of observation. Currently we don't have clear explanation for this strange spectrum, although it can be a relatively long period, high eccentric binary, observed at the moment of periastron passage, so $\drv$ was significant for detection.    

\subsection{Gaia DR3 data for LAMOST-MRS spectra}
\begin{figure}
	\includegraphics[width=\columnwidth]{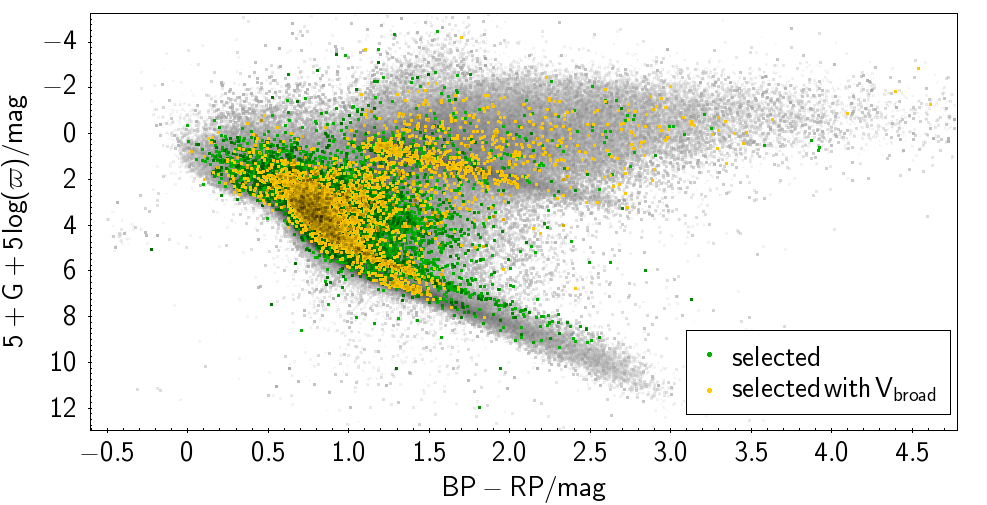}
    \includegraphics[width=\columnwidth]{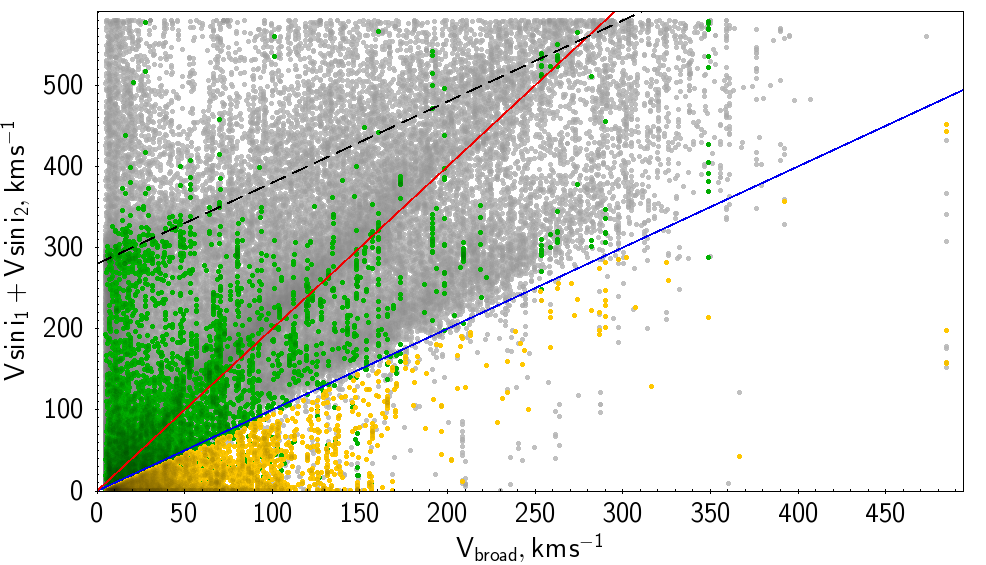}
    \caption{Gaia DR3 Hertzsprung-Russell diagram for observed LAMOST-MRS dataset (top panel) and $\vsini_1+\vsini_2$ versus $v_{\rm broad}$ (bottom panel). SB2 candidates selected in Fig.\protect\ref{fig:select} are shown with green circles, SB2 candidates selected using $v_{\rm broad}$ are shown with yellow circles.}
    \label{fig:hrd}
\end{figure}

%583 609 matches
We check the recent Gaia DR3 \citep[][]{gaia_all,gaia3} and plot the Hertzsprung-Russell diagram for all matches with positive parallax in top panel of Fig~\ref{fig:hrd}. 
Selected SB2 candidates are shown with green triangles. Some of them are located at the main sequence of binaries with similar luminosity which is $\sim0.75$ mag higher than main sequence. Gaia DR3 provides parameter $v_{\rm broad}$ as a measure of rotational broadening for a single stars, based on Gaia RVS spectra \citep[][]{gaia_vbroad}. We use it instead of $\vsini_0$ and reproduce bottom panel of Fig.~\ref{fig:select} in the bottom panel of Fig.~\ref{fig:hrd}.  Similarly to Fig.~\ref{fig:select}, not-selected stars slightly follow the red and blue lines. We have an overdensity at $v_{\rm broad}<10\,\kms$, and we have many hot stars in the region higher than the black dashed line $\vsini_1+\vsini_2=v_{\rm broad}+300$. In Fig.~\ref{fig:select} this space was empty, because our fitting algorithm is unable fit $\teff>8800$~K, so it tries to increase $\vsini_0$ in order to compensate for changes in the spectra. We apply the same selection as before using $v_{\rm broad}$ instead of $\vsini_0$ and find 6116 spectra (3764 stars), where 3784 (1880 stars) of them were previously selected using $\vsini_0$. We show them as green and yellow circles respectively. The remaining stars were probably observed by LAMOST-MRS at moments with $\drv\sim0~\kms$. Unfortunately we can't check this hypothesis, because epoch's RVS spectra and RV measurements aren't included in Gaia DR3.% However this confirms that \cite{gaia_vbroad} were unable to completely filter out SB2s from their catalogue. 

\par
\begin{figure}%[!htb]
	\includegraphics[width=\columnwidth]{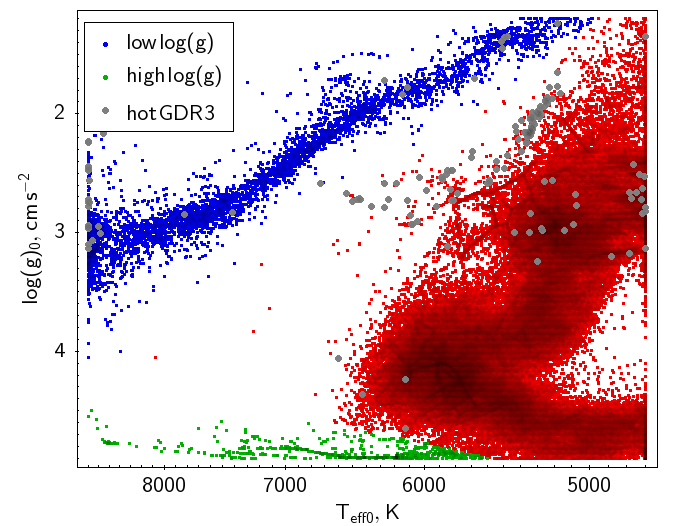}
    \includegraphics[width=\columnwidth]{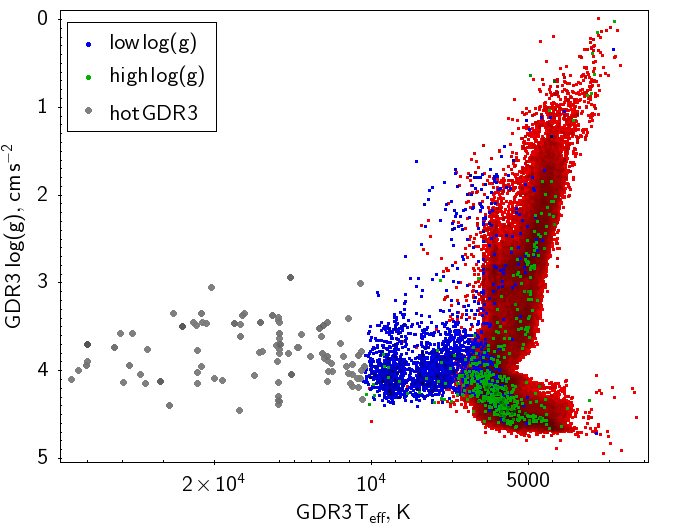}
    \caption{Comparison of the Kiel diagram for selected subset of the single stars (top panel) with corresponding Kiel diagram for their matches in Gaia DR3 (bottom panel).  }
    \label{fig:singles}
\end{figure}

We compare our estimates for the subset of single stars with good fit spectra (selected using cuts in Table~\ref{tab:cuts}) and $\chi_{\rm single} < \chi_{\rm binary}$ with their matches in Gaia DR3 estimates in Figure~\ref{fig:singles}. It is evident that the sequence of stars with $\logg_0<3$ have $\logg\sim4$ in Gaia DR3 (shown as blue markers). In \cite{j0647} spectroscopic $\logg$ were also underestimated in comparison with values derived during the modelling of light curves, although the bias was much smaller. Thus we think that information in LAMOST-MRS spectra alone can not confidently constrain the surface gravity for stars with $\teff>6500$ K. Another group with high $\logg_0>4.5$ dex and $\teff_0>5700$ K, shown as green circles, in Gaia DR3 have $\logg$ in range from 0 to 4.6 dex. We checked their spectral plots and eventually almost all of them don't have spectral data in red arm. Hot stars in Gaia DR3 with $\teff>10000$ K were mostly placed to the left of the red giants branch. Inspection of their plots found that many of them have no spectral features except strong emission in $\ha$. Other such stars were placed at the maximal value $\teff\sim8790$ K or at the red giant branch. These red giants were confirmed by the visual inspection of the plots and probably were misclassified by Gaia DR3.    %189060 matches with params, of 254993 singles  

The estimations of the spectral parameters can have significant biases. These parameters should be seen not as exact measurements, but more like set of parameters that represent best fit flexible template, similar to one used for RV determination using cross-correlation functions (see \cite{gaia3} that have published best template parameters for radial velocity spectrograph).

\section{Conclusions}
\label{concl}
We have used an updated method for the detection of double-lined spectroscopic binaries (SB2) using $\vsini$ values from spectral fits. The method is applied to all spectra from LAMOST-MRS. This method found 12426 (8105 new) SB2 candidates in the LAMOST-MRS. The updated method allows one to select many SB2 candidates, including ones with zero $\Delta\rv$ between components, if they have different spectra. We present a determination of the mass ratios and orbits for a subset of SB2 candidates with multiple observations. We also detect several cases where the spectra are contaminated by solar light, which can be identified by using our binary spectral model. % We hope that our method will be useful in SB2 detection in large-scale spectroscopic surveys, e.g. Gaia RVS \citep{grvs}.  %(similar to previous works \citep{tyc,j0647})
 
%The last numbered section should briefly summarise what has been done, and describe
%the final conclusions which the authors draw from their work.

\section*{Acknowledgements}
 We are grateful to the anonymous referee for a constructive report. We thank Hans B{\"a}hr for his careful proof-reading of the manuscript.
%MK is grateful to his parents, Yuri Kovalev and Yulia Kovaleva, for their full support in making this research possible. MK thanks Maria Kovaleva for valuable discussions.  
This work is supported by National Key R\&D Program of China (Grant No. 2021YFA1600401/3), and by the Natural Science Foundation of China (Nos. 12090040/3, 12125303, 12288102, 11733008).
Guoshoujing Telescope (the Large Sky Area Multi-Object Fiber Spectroscopic Telescope LAMOST) is a National Major Scientific Project built by the Chinese Academy of Sciences. Funding for the project has been provided by the National Development and Reform Commission. LAMOST is operated and managed by the National Astronomical Observatories, Chinese Academy of Sciences. The authors gratefully acknowledge the “PHOENIX Supercomputing Platform” jointly operated by the Binary Population Synthesis Group and the Stellar Astrophysics Group at Yunnan Observatories, Chinese Academy of Sciences. 
This work has made use of data from the European Space Agency (ESA) mission {\it Gaia} (\url{https://www.cosmos.esa.int/gaia}), processed by the {\it Gaia} Data Processing and Analysis Consortium (DPAC, \url{https://www.cosmos.esa.int/web/gaia/dpac/consortium}). Funding for the DPAC has been provided by national institutions, in particular the institutions participating in the {\it Gaia} Multilateral Agreement.
%Based on data products from observations made with ESO Telescopes at the La Silla Paranal Observatory under run IDs 188.B-3002 and 193.B-0936.
This research has made use of NASA’s Astrophysics Data System, the SIMBAD data base, and the VizieR catalogue access tool, operated at CDS, Strasbourg, France. It also made use of TOPCAT, an interactive graphical viewer and editor for tabular data \citep[][]{topcat}.

\section*{Data Availability}
The data underlying this article will be shared on reasonable request to the corresponding author.
LAMOST-MRS spectra are downloaded from \url{www.lamost.org}.

%%%%%%%%%%%%%%%%%%%%%%%%%%%%%%%%%%%%%%%%%%%%%%%%%%



\bibliographystyle{mnras}

%\begin{thebibliography}{99}
%\end{thebibliography}

%%%%%%%%%%%%%%%%%%%%%%%%%%%%%%%%%%%%%%%%%%%%%%%%%%

%%%%%%%%%%%%%%%%% APPENDICES %%%%%%%%%%%%%%%%%%%%%

\appendix

\section{Spectral model for LAMOST-MRS}
\label{sec:payne}
\begin{figure*}%[!htb]
	\includegraphics[width=0.83\textwidth]{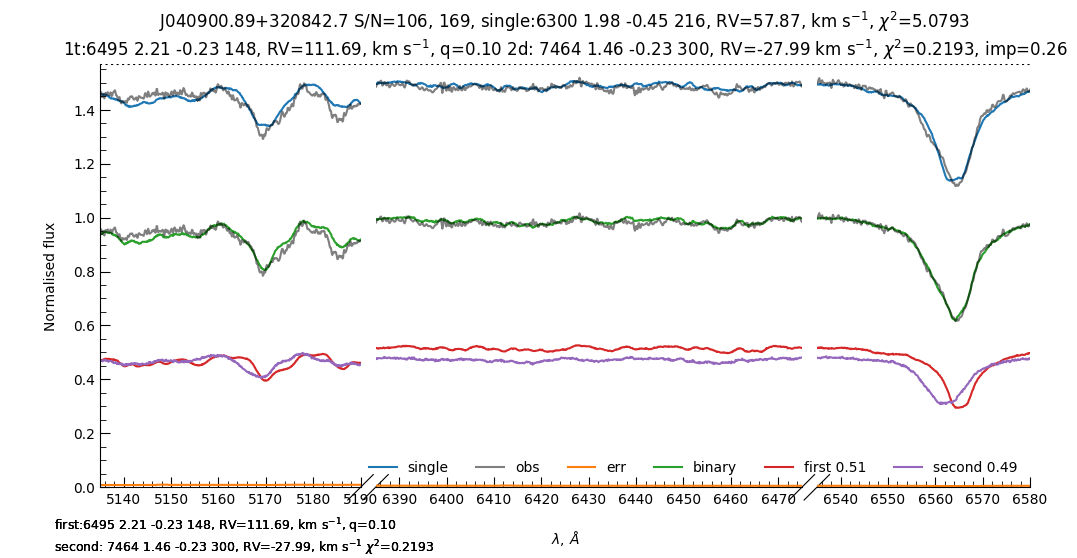}
	\includegraphics[width=0.83\textwidth]{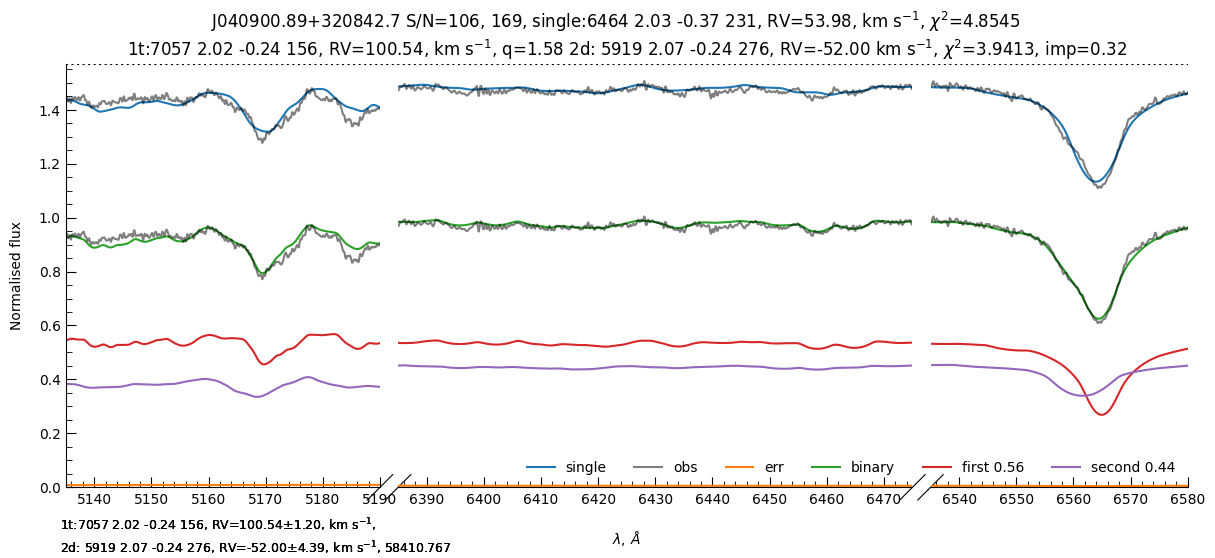}
    \caption{Same as Fig.\protect\ref{fig:moon} but with results SB2 candidate of two fast rotators fitted with NN generated spectral model (top) and with linear interpolation of spectral models (bottom).}% Observed spectrum and its error are shown as a gray and orange lines respectively. Primary (magenta line) and secondary (red line) components are labeled as "second" and "first" with contribution to total light at $\lambda=5000$~\AA.~Spectral parameters ($\teff,~\logg,~\feh,~\vsini$) from single star model fit and binary model fit are shown in the titles.}
    \label{fig:artefact}
\end{figure*}

The grid of synthetic spectra (6200 in total) is generated using the NLTE~MPIA online-interface \url{https://nlte.mpia.de} \citep[see Chapter~4 in][]{disser} on wavelength intervals 4870:5430 \AA~for the blue arm and 6200:6900 \AA ~for the red arm with spectral resolution $R=7500$. We use a NLTE (non-local thermodynamic equilibrium) spectral synthesis for H, Mg~I, Si~I, Ca~I, Ti~I, Fe~I and Fe~II lines \citep[see Chapter~4 in][ for references]{disser}.  %876 nlte lines blue, 378 red
The spectral parameters are randomly selected in a range of $\teff$=4600, 8800 K, $\logg$=1.0, 4.99 (cgs units), $\vsini$= 0, 300 $\kms$ and [Fe/H]\footnote{We used $\feh$ as a proxy of overall metallicity, abundances for all elements are scaled with Fe.}=$-$0.9,$+$0.9 dex,
microturbulence is fixed to $\Vmic=2~\kms$. %The grid is randomly split on training (70\%) and cross-validation (30\%) sets of spectra, which are used to train \textit{The~Payne} spectral model \citep{ting2019}. %The neural network (NN) consists of two layers of 300 neurons each with rectilinear unit (ReLU)\footnote{ReLU(x)=max(x,0)} activation functions. We train separate NNs for each spectral arm. The median approximation error is less than 1\% for both arms. 
We found that neural network (NN) generated spectrum used in \cite{cat22} can contain small artefacts (usually smaller or comparable to the noise level) that disturb correct rotational profiles for a broad lines, which can lead to errors. Thus we replaced NN with a standard linear interpolation in four-dimensional space and used output of \texttt{scipy.interpolate.LinearNDinterpolate} as a single-star spectral model ${f}_{\lambda,{\rm single}}$.  See Figure~\ref{fig:artefact} with comparison of NN and linear interpolation results for one spectrum. It is clear that a linear interpolation provides smooth line profiles with ``bell-like" shape for $\ha$, while line profile generated by NN has the wrong shape.

\section{Catalogue of SB2 candidates in LAMOST-MRS}

Table~\ref{tab:final} lists all 12426 SB2 candidates. We note that LAMOST designation can slightly vary between data releases, therefore we recommend to use Gaia DR3 \texttt{source\_id} from \citet{gaia3}.  

\begin{table}%\small
    \centering
    \caption{Catalogue of SB2 candidates in LAMOST-MRS. Full table is available as supplementary material.}
    \begin{tabular}{cc}
\hline
\hline
LAMOST designation & Gaia DR3 \texttt{source\_id} \\
\hline

J000000.18+320847.1 & 2873824744457023360\\
J000006.63+492648.6 & 393492950765727872\\
J000021.70+383903.8 & 2880989157927180800\\
..&..\\

\hline
    \end{tabular}
    \label{tab:final}
\end{table}

\section{Data tables}

In Tables~\ref{tab:wilson},\ref{tab:orbits} we present results for detailed analysis of SB2 candidates with multiple spectra.

\begin{table*}%\small
    \centering
    \caption{Dynamical mass ratios and systemic velocities. Full table is available as supplementary material.}
    \begin{tabular}{cccc}
\hline
\hline

Star & $\gamma_{\rm dyn}$, $\kms$ & $Q_{\rm dyn}$ &   Gaia DR3 \texttt{source\_id}  \\ 
\hline
J000223.56+340644.8 & $-38.84\pm0.16$ & $1.098\pm0.006$ & 2875159070536814464\\
J000247.24+351645.5 & $-15.36\pm0.98$ & $1.790\pm0.057$ & 2876964296830796160\\
J000429.80+363651.3 & $-16.51\pm0.47$ & $1.04\pm0.02$ & 2877196774820732288\\
..& .. &.. & ..\\

\hline
    \end{tabular}
    \label{tab:wilson}
\end{table*}

\begin{table}%\small
    \centering
    \caption{Orbital parameters from {\sc GLS}. Full table is available as supplementary material.}
    \begin{tabular}{cc}
\hline
\hline

% star & &  ind. epoch & single  \\
parameter & unit \\

\hline
NAME & JHHMMSS.SS+DDMMSS.S\\
$N_{\rm sp}$ & \\%, Psin, ePsin, Ksin, eKsin, T0sin, eT0sin, Csin, eCsin, Pkep, ePkep, K, eK, RV0, eRV0, T0, eT0, ecc, omega 
$P_{\rm sin}$ & d\\
$\sigma P_{\rm sin}$ & d\\
$K_{\rm sin}$ & $\kms$\\
$\sigma K_{\rm sin}$ & $\kms$\\
${t_0}_{\rm sin}$ & d\\
$\sigma {t_0}_{\rm sin}$ & d\\
$\gamma_{\rm sin}$ & $\kms$\\
$\sigma \gamma_{\rm sin}$ & $\kms$\\
$P_{\rm kep}$ & d\\
$\sigma P_{\rm kep}$ & d\\
$K_{\rm kep}$ & $\kms$\\
$\sigma K_{\rm kep}$ & $\kms$\\
${t_0}_{\rm kep}$ & d\\
$\sigma {t_0}_{\rm kep}$ & d\\
$\gamma_{\rm kep}$ & $\kms$\\
$\sigma \gamma_{\rm kep}$ & $\kms$\\
$e$ &\\
$\omega$ & deg\\
  Gaia DR3 \texttt{source\_id} & \\
\hline
    \end{tabular}
    \label{tab:orbits}
\end{table}

%%%%%%%%%%%%%%%%%%%%%%%%%%%%%%%%%%%%%%%%%%%%%%%%%%

% Don't change these lines
\bsp	% typesetting comment
\label{lastpage}
\end{document}